\documentstyle[psfig,twocolumn,amssymb,pre,aps]{revtex}
\newcommand{\alo}{\mbox{${\rm Al}_2{\rm O}_3$}}

\begin{document}

\wideabs{
\title{
Partial Resonances of Three-Phase Composites at Long Wavelengths }

\author{N. A. Nicorovici and R. C. McPhedran}
\address{School of Physics, University of Sydney, NSW 2006, Australia}
\author{G. W. Milton}
\address{Department of Mathematics, University of Utah,
Salt Lake City, UT 84112, USA}
\author{L. C. Botten}
\address{Department of Mathematical Sciences, University of
Technology, Sydney, Broadway, P O Box 123, NSW 2007, Australia\\~\\}

\date{\today}
\maketitle

\begin{abstract}
We investigate the behaviour of a three-phase composite,
structured as a hexagonal array of coated cylinders, when the sum
of the dielectric constants of the core and shell equals zero. In
such cases, the absorption of the composite is the same as the
absorption of a periodic array of solid cylinders of core material
and radius equal to the outer radius of the original coated
cylinder. When the sum of the dielectric constants of the shell
and matrix equals zero, the composite has the same absorption as a
periodic array of solid cylinders of core material, and radius
exceeding the shell radius.
%We show that this effect is also
%related with the anomalous absorptance of metallic cylinders.
\end{abstract}
}

\section{Introduction}
\label{intro}

Previous studies of the effective dielectric constant of a
periodic array of coated cylinders, in the quasistatic limit, have
revealed some unexpected and interesting
results\cite{cc1,cc2,cc3}.
Specifically, we have used the method of Lord
Rayleigh\cite{ray} to investigate the analytic properties of the
effective dielectric constant of the array, as a function of the core
dielectric constant ($\varepsilon_c$) and the shell dielectric constant
($\varepsilon_s$), while keeping the matrix dielectric constant
($\varepsilon_m$) fixed.

We have shown that when
$\varepsilon_c+\varepsilon_s=0$, the composite has exactly the
same effective dielectric constant as a periodic array of solid
cylinders with dielectric constant
$\varepsilon_c$ and radius equal to the
outer radius of the original coated cylinder. We have also shown
that when
$\varepsilon_s+\varepsilon_m=0$,
the composite has exactly the same effective
dielectric constant as a periodic array of solid cylinders with
dielectric constant $\varepsilon_c$,
and radius exceeding the shell radius.
In both cases, the core is {\em magnified} and it may be possible for
a system with a vanishingly small concentration
to exhibit properties like those of a concentrated system.
This is what we have called a {\em partially resonant} system,
since the system has
a finite, rather than an infinite, response to the applied field.

The same effect appears in the case of a random mixture
of coated cylinders with nonlinear cores and linear
shells\cite{ohad}.
Under conditions of partial resonance, the system behaves as a
mixture of nonlinear solid cylinders of radii equal to, or
larger than, those of the original coated cylinders.

\begin{figure}
\vspace{6ex}
\centerline{\psfig{file=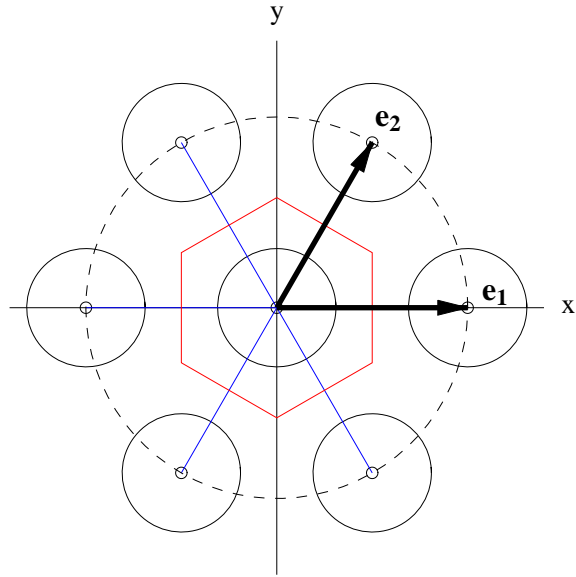,height=3.0in}}
\vspace{6ex}
\caption{The Wigner-Seitz cell (red) with the central inclusion
(coated cylinder with $r_c=0.03d$, $r_s=0.31d$ and $d=1$) and the
nearest neighbours (located along the dashed circle) for a
hexagonal array. The centers of the nearest neighbours are given
by the lattice vectors ${\bf R}_p$ for
$p\in\{(-1,0),(-1,1),(0,-1),(0,1),(1,-1),(1,0)\}$. We also show
the fundamental translation vectors ${\bf e}_1=d(1, 0)$ and  ${\bf
e}_2=d(1/2, \sqrt{3}/2)$.} \label{WSC}
\end{figure}

Here, we show how these results carry over to problems of
electromagnetic scattering by periodic arrays of
coated cylinders, and elucidate the relation between
partial resonances and
anomalous absorptance of metallic wires \cite{anomaly}.
From the many possible ways of packing cylinders in
regular arrays in two dimensions we concentrate on the hexagonal
lattice that is isotropic for fields applied in the plane
perpendicular to the cylinder axes (see Fig.~\ref{WSC}).
However, the method we use
may be applied equally well to any periodic
array of coated cylinders.

%%%%%%%%%%%%%%%%%%%%%%%%%%%%%%%%%%%%%%%%%%%%%%%%%%%%%%%%%%%%%%%%%%%%%%%
%%%%%%%%%%%%%%%%%%%%%%%%%%%%%%%%%%%%%%%%%%%%%%%%%%%%%%%%%%%%%%%%%%%%%%%

Recent work by Pendry {\em et al}. \cite{pendry} has heightened interest in the
extraordinary and sometimes paradoxical behaviour which can arise
in composite systems, in which one component has a dielectric constant
(or a magnetic permeability) which takes approximately a real and
negative value.
In such a composite we may observe strong focusing of light at a plane
interface, or ultra-refraction \cite{gralak}, or a dilute composite
{\em masquerading} as a concentrated one \cite{cc1,cc2,cc3}.
Again, the use of metal-coated spheres in photonic crystals has resulted
in robust band gaps, independent of sphere-packing geometries \cite{pingsheng}.

\section{The Field Identity}
\label{raysec}
\subsection{Field Expansions}
\label{fieldssec}

We consider the electromagnetic modes of a three--phase composite
consisting of an array of coated cylinders of infinite length, each
aligned parallel to the $z$ axis,
inserted into a matrix of dielectric constant $\varepsilon_m$ and
magnetic permeability $\mu_m$.
For each cylinder, the core and shell are characterized by the
radii $r_c$ and  $r_s$, dielectric constants
$\varepsilon_c$ and $\varepsilon_s$, and magnetic permeabilities
$\mu_c$ and $\mu_s$, respectively.
We also assume that the electric and magnetic fields depend on
time through a factor $\exp{(- i \omega t)}$.
The core and shell of all cylinders, as well as the matrix are
homogeneous so that, the fields satisfy the Maxwell equations:
\begin{eqnarray}
\nabla \times {\bf H} = & - & i \omega \varepsilon({\bf r}) {\bf E} ~ , ~ ~
\nabla \cdot {\bf E} = 0 ~ ,
\label{a1} \\
\nabla \times {\bf E} = &   & i \omega \mu({\bf r}) {\bf H} ~ , ~ ~
\nabla \cdot {\bf H} = 0 ~ ,
\label{a2}
\end{eqnarray}
with
\begin{equation}
\varepsilon({\bf r}) = \left\{
\begin{array}{lcl}
\varepsilon_m  & ~~ {\rm for} ~~ & {\bf r} \in \Omega_m \, , \\
\varepsilon_s  & ~~ {\rm for} ~~ & {\bf r} \in \Omega_s \, , \\
\varepsilon_c  & ~~ {\rm for} ~~ & {\bf r} \in \Omega_c \, , \\
\end{array} \right.
\label{a3}
\end{equation}
and
\begin{equation}
\mu({\bf r}) = \left\{
\begin{array}{lcl}
\mu_m  & ~~ {\rm for} ~~ & {\bf r} \in \Omega_m \, , \\
\mu_s  & ~~ {\rm for} ~~ & {\bf r} \in \Omega_s \, , \\
\mu_c  & ~~ {\rm for} ~~ & {\bf r} \in \Omega_c \, . \\
\end{array} \right.
\label{a4}
\end{equation}
Here, in the $xy-$plane, $\Omega_c$ denotes the union of all the
domains occupied by cylinder cores, while $\Omega_s$
denotes the union of all the domains occupied by the cylinder shells.
Thus, $\Omega_c \bigcup \Omega_s$ represents the region occupied by
all the cylinders in the array and $\Omega_m$ is the region between
the cylinders (the matrix region).

By eliminating ${\bf E}$ or ${\bf H}$ between (\ref{a1}) and (\ref{a2}),
we obtain the Helmholtz equations for the components of the fields:
\begin{equation}
\left( \nabla^2 + \kappa^2 \right)
\left\{ \begin{array}{c} {\bf E} \\ {\bf H} \end{array} \right\} = 0
\, ,
\label{a5}
\end{equation}
where $\kappa^2=\omega^2 \varepsilon \mu$ is again a function of
position.
The axes of the cylinders in the array are parallel to the $z-$axis, and
the fields are taken to depend on $z$ through the factor $\exp(i \beta z)$,
with $\beta$ representing the propagation constant along the $z$
direction, in all the regions $\Omega_c$, $\Omega_s$ and $\Omega_m$.
Also, in cylindrical coordinates ($r$, $\theta$, $z$), the field
components $E_z$ and $H_z$ determine the other four field components
through the relations\cite{simon}:
\begin{eqnarray}
E_r & = &
\frac{i}{\kappa^2 - \beta^2}
\left[ \beta \, \frac{\partial E_z}{\partial r} +
\frac{\omega \mu}{r} \, \frac{\partial H_z}{\partial \theta}
\right]
% =
%\frac{i}{\kappa^2 - \beta^2}
%\left[ \beta \, \frac{\partial E_z}{\partial \nu} +
%\omega \mu \, \frac{\partial H_z}{\partial \tau}
%\right]
\, ,
\label{a7} \\
E_\theta & = &
\frac{i}{\kappa^2 - \beta^2}
\left[ \frac{\beta}{r} \, \frac{\partial E_z}{\partial \theta} -
\omega \mu \, \frac{\partial H_z}{\partial r} \right]
% =
%\frac{i}{\kappa^2 - \beta^2}
%\left[ \beta \, \frac{\partial E_z}{\partial \tau} -
%\omega \mu \, \frac{\partial H_z}{\partial \nu} \right]
\, ,
\label{a8} \\
H_r & = &
\frac{- i}{\kappa^2 - \beta^2}
\left[ \frac{\omega \varepsilon}{r} \,
\frac{\partial E_z}{\partial \theta} -
\beta \, \frac{\partial H_z}{\partial r}
\right]
%=
%\frac{- i}{\kappa^2 - \beta^2}
%\left[ \omega \varepsilon \,
%\frac{\partial E_z}{\partial \tau} -
%\beta \, \frac{\partial H_z}{\partial \nu}
%\right]
\, ,
\label{a9} \\
H_\theta & = &
\frac{i}{\kappa^2 - \beta^2}
\left[ \omega \varepsilon \, \frac{\partial E_z}{\partial r} +
\frac{\beta}{r} \, \frac{\partial H_z}{\partial \theta} \right]
% =
%\frac{i}{\kappa^2 - \beta^2}
%\left[ \omega \varepsilon \, \frac{\partial E_z}{\partial \nu} +
%\beta \, \frac{\partial H_z}{\partial \tau} \right]
\, .
\label{a10}
\end{eqnarray}
The field components $E_z$ and $H_z$ satisfy the Helmholtz equations:
\begin{equation}
\left( \nabla_\perp^2 + k^2 \right)
\left\{ \begin{array}{c} E_z \\ H_z \end{array} \right\} = 0 \, ,
\label{a11}
\end{equation}
where $\nabla_\perp^2$ is the transverse Laplacian and
\begin{equation}
k^2 = \kappa^2 - \beta^ 2 \, ,
\label{a12}
\end{equation}
represents the transverse (in-plane) propagation constant.

For the cylinder centered at the origin
of coordinates, we represent the electric and magnetic fields
$E_z$ and $H_z$ (denoted here by $V$),
by series expansions in terms of cylindrical harmonics:
\begin{equation}
V(r, \theta) =
\left\{
\begin{array}{l}
{\displaystyle \sum_{l=-\infty}^\infty A_l^{c} J_l(k_c r)
 \, e^{i l \theta}}  \, ,
\\
% && \\
{\displaystyle \sum_{l=-\infty}^\infty
\left[ A_l^{s} J_l(k_s r) +
 B_l^{s} Y_l(k_s r) \right]  e^{i l \theta}}\,  ,
\\
% && \\
{\displaystyle \sum_{l=-\infty}^\infty
\left[ A_l^{m} J_l(k_m r) +
 B_l^{m} Y_l(k_m r) \right]  e^{i l \theta}} \, ,
\end{array}
\right.
\label{fields}
\end{equation}
where $J$ and $Y$ represent the Bessel functions of the first
and second kind. The three forms of the series expansions
in (\ref{fields}) correspond to
$0 \leq r \leq r_c$ (i.e., inside the core),
$r_c \leq r \leq r_s$ (i.e., inside the shell) and
$r \geq r_s$ (i.e., in the matrix), respectively.
Also, the superscripts $c$, $s$ and $m$ label the fields inside the
cylinder core, cylinder shell, and in the matrix, respectively.
Thus, we have $k_c^2=\omega^2 \varepsilon_c \mu_c$,
$k_s^2=\omega^2 \varepsilon_s \mu_s$, and
$k_m^2=\omega^2 \varepsilon_m \mu_m$.

\subsection{Boundary Conditions}
\label{bcdsec}

The boundary conditions express
the continuity of the tangential components of the electric
($E_z$, $E_\theta$)
and magnetic ($H_z$ and $H_\theta$) fields across the core
and shell surfaces.
When the crystal momentum is perpendicular
to the axes of the cylinders in the array, we have $\beta=0$,
and the problem can be reduced to solving two independent
problems\cite{panofsky}:
(i) for TM or $s$ polarization (when $H_z=0$ and the transverse
parts of ${\bf E}$ and ${\bf H}$ are generated by $\nabla_\perp E_z$), and
(ii) for TE or $p$ polarization (when $E_z=0$ and
$\nabla_\perp H_z$ gives the transverse components of ${\bf E}$ and
${\bf H}$).
Thus, depending on the polarization, we obtain:
\begin{eqnarray}
\left. E_z^{(c)} \right|_{r=r_c} & = & \left. E_z^{(s)}
\right|_{r=r_c},
\label{a13x} \\
\left. \frac{1}{\mu_c} \, \frac{\partial E_z^{(c)}}{\partial r}
\right|_{r=r_c}
 & = &
\left. \frac{1}{\mu_s} \, \frac{\partial E_z^{(s)}}{\partial r}
\right|_{r=r_c} ,
\label{a14x} \\
\left. E_z^{(s)} \right|_{r=r_s} & = & \left. E_z^{(m)}
\right|_{r=r_s},
\label{a13} \\
\left. \frac{1}{\mu_s} \, \frac{\partial E_z^{(s)}}{\partial r}
\right|_{r=r_s}
 & = &
\left. \frac{1}{\mu_m} \, \frac{\partial E_z^{(m)}}{\partial r}
\right|_{r=r_s}  ,
\label{a14}
\end{eqnarray}
for TM polarization, and:
\begin{eqnarray}
\left. H_z^{(c)} \right|_{r=r_c} & = & \left. H_z^{(s)}
\right|_{r=r_c},
\label{a15x} \\
\left. \frac{1}{\varepsilon_c} \, \frac{\partial H_z^{(c)}}{\partial r}
\right|_{r=r_c}
 & = &
\left. \frac{1}{\varepsilon_s} \, \frac{\partial H_z^{(s)}}{\partial r}
\right|_{r=r_c} ,
\label{a16x} \\
\left. H_z^{(s)} \right|_{r=r_s} & = & \left. H_z^{(m)}
\right|_{r=r_s},
\label{a15} \\
\left. \frac{1}{\varepsilon_s} \, \frac{\partial H_z^{(s)}}{\partial r}
\right|_{r=r_s}
 & = &
\left. \frac{1}{\varepsilon_m} \, \frac{\partial H_z^{(m)}}{\partial r}
\right|_{r=r_s}  ,
\label{a16}
\end{eqnarray}
for TE polarization.

In the case of TM polarization, for the cylinder centered at the origin
of coordinates, we use the series expansions (\ref{fields}) for
$E_z$, and
by means of the boundary conditions (\ref{a13x}-\ref{a14})
we obtain relations of the form:
\begin{eqnarray}
\left[ \begin{array}{c} A_l^m \\ B_l^m \end{array} \right]
& = &
\left[ \begin{array}{rr}
J_l(k_m r_s) & Y_l(k_m r_s) \\
Z_m^{-1} J_l'(k_m r_s) & Z_m^{-1} Y_l'(k_m r_s) \end{array}
\right]^{-1}
\nonumber \\ & \times &
\left[ \begin{array}{rr}
J_l(k_s r_s) & Y_l(k_s r_s) \\
Z_s^{-1} J_l'(k_s r_s) & Z_s^{-1} Y_l'(k_s r_s) \end{array}
\right]
 \label{TMbcd} \\ & \times &
\left[ \begin{array}{rr}
J_l(k_s r_c) & Y_l(k_s r_c) \\
Z_s^{-1} J_l'(k_s r_c) & Z_s^{-1} Y_l'(k_s r_c) \end{array}
\right]^{-1}
\nonumber \\ & \times &
\left[ \begin{array}{rr}
J_l(k_c r_c) & Y_l(k_c r_c) \\
Z_c^{-1} J_l'(k_c r_c) & Z_c^{-1} Y_l'(k_c r_c) \end{array}
\right]
\left[ \begin{array}{c} A_l^c \\ 0 \end{array} \right]
\, , \nonumber
\end{eqnarray}
or
\begin{equation}
A_l^m= - M_l  B_l^m \, .
\label{a21}
\end{equation}
In (\ref{TMbcd}), the prime indicates the derivative of the
corresponding function and $Z_i=\sqrt{\mu_i/\varepsilon_i}$
($i = m, s, c$) represent the impedances of the matrix, shell
and core, respectively.
For TE polarization we obtain the relation between $A_l^m$
and $B_l^m$ by changing $Z_i\rightarrow1/Z_i$ in (\ref{TMbcd}).

The structure of (\ref{TMbcd}) allows us a generalization to
multicoated cylinders, and for this reason we have added the
second column in the last matrix. However, this last matrix is
multiplied by the vector $[A_l^c, 0]^T$ corresponding to the field
expansion inside the core (i.e., the first form in
(\ref{fields})).
Note that the first two matrices in the right side of
(\ref{TMbcd}) arise from the boundary conditions at the interface
between the shell and matrix ((\ref{a13}) and
(\ref{a14})), while the next two matrices come from the boundary
conditions at the core -- shell interface ((\ref{a13x}) and
(\ref{a14x})). Consequently, in the case of a multicoated cylinder,
we may infer the form of a general term representing the interface
$i$ (between the $i^{\rm th}$ and $(i+1)^{\rm th}$ media)
characterized by the radius $r_i$:
\begin{eqnarray}
{\bf M}_l^{i, i+1} & = &
\left[ \begin{array}{rr}
J_l(k_{i} r_i) & Y_l(k_{i} r_i) \\
Z_{i}^{-1} J_l'(k_{i} r_i) & Z_{i}^{-1} Y_l'(k_{i} r_i) \end{array}
\right]^{-1}
\nonumber \\ & \times &
\left[ \begin{array}{rr}
J_l(k_{i+1} r_i) & Y_l(k_{i+1} r_i) \\
Z_{i+1}^{-1} J_l'(k_{i+1} r_i) & Z_{i+1}^{-1} Y_l'(k_{i+1} r_i) \end{array}
\right] .
\label{bcdterm}
\end{eqnarray}
Here, we have counted the layers from the exterior to interior and
thus, for $n-1$ shells, the matrix is indexed by $i=0$ and the core
(the innermost medium) by $i=n$. Now, Eq.~(\ref{TMbcd}) takes the form
\begin{equation}
\left[ \begin{array}{c} A_l^0 \\ B_l^0 \end{array} \right]
 =
\left( \prod_{i=0}^{n-1} {\bf M}_l^{i, i+1} \right)
\left[ \begin{array}{c} A_l^{n+1} \\ 0 \end{array} \right] \, .
\label{TMbcdgen}
\end{equation}
We also obtain the expression for the TE polarization by
substituting $Z_i\rightarrow1/Z_i$.

\subsection{The Generalized Rayleigh Identity}
\label{GRIsec}

For each polarization (TM or TE), we may write
a Generalized Rayleigh Identity of the form\cite{cnr,nrb1}:
\begin{equation}
M_l B_l
  +  \sum_{n=-\infty}^\infty
S_{l-n}^Y(k_m, {\bf k}_0) \, B_n = 0 \, ,
\label{gri}
\end{equation}
with $M_l$ from (\ref{a21}) and $-\infty < l < \infty$.
The quantities $S_l^Y$ are dynamic lattice sums, which may be evaluated
using accelerated summation over the reciprocal array\cite{cnr}:
\begin{eqnarray}
&& S_l^Y(k_m, {\bf k}_0) J_{l+n}(k_m \xi) =
\nonumber \\
&& -
\left[ Y_n(k_m \xi) + \frac{1}{\pi} \sum_{q=1}^n
\frac{(n-q)!}{(q-1)!} \left( \frac{2}{k_m \xi} \right)^{n - 2 q +2} \right]
\delta_{l,0} \nonumber \\
&& -  i^l \frac{4}{A} \sum_{h} \left( \frac{k_m}{Q_h} \right)^n
\frac{J_{l+n}(Q_h \xi)}{Q_h^2 - k_m^2}
e^{i l \theta_h} .
\label{rls}
\end{eqnarray}
Here, $n$ is an arbitrary non--negative integer, $\xi$ is the length of
an arbitrary vector inside the unit cell (shorter than the shortest
line connecting two of its vertices), and
$A$ represents the area of the unit cell. The vector ${\bf k}_0$
is the in--plane component of the ``crystal momentum''
${\bf k}_B = {\bf k}_0 + \beta {\bf {\widehat{z}}}$,
with ${\bf {\widehat{z}}}$ the unit vector along the $z-$axis.
Note that, in this paper we consider only the case $\beta=0$,
so that ${\bf k}_B = {\bf k}_0$.

The quantities ${\bf Q}_h$ and $\theta_h$ from (\ref{rls}) depend
on the type of the array. In our case, the hexagonal array is
defined by the fundamental translation vectors ${\bf e}_1=d(1, 0)$
and ${\bf e}_2=d(1/2, \sqrt{3}/2)$, where $d$ is the array
constant (see Fig.~\ref{WSC}). Hence, the centers of the cylinders
are located at the points given by the radius vectors
\begin{equation}
{\bf R}_p = p_1 {\bf e}_1 + p_2 {\bf e}_2 ,
\end{equation}
with $p$ denoting the pair of integers
$(p_1, p_2)\in{\mathbb Z}^2$.
The fundamental vectors of the reciprocal array are given by the
relations
\begin{eqnarray}
{\bf u}_1 & = &
\frac{2\pi}{A}{\bf e}_2 \times {\bf e}_3
= \frac{2\pi}{d} \left( 1, - \frac{1}{\sqrt{3}} \right) ,
\label{u1} \\
{\bf u}_2 & = &
\frac{2\pi}{A} {\bf e}_3 \times {\bf e}_1
= \frac{2\pi}{d} \left( 0,  \frac{2}{\sqrt{3}} \right) ,
\label{u2}
\end{eqnarray}
where $A=\|{\bf e}_1 \times {\bf e}_2\|$
and
${\bf e}_3=({\bf e}_1 \times {\bf e}_2)/A$
is a dimensionless unit vector, along the $z$ axis.
With the vectors (\ref{u1}) and (\ref{u2}) we form the
reciprocal lattice vectors
\begin{equation}
{\bf K}_h = h_1 {\bf u}_1 + h_2 {\bf u}_2 ,
\end{equation}
where $h=(h_1,h_2)\in{\mathbb Z}^2$.
Finally, we have
${\bf Q}_h = {\bf k}_0 + {\bf K}_h$
and $\theta_h = \arg{({\bf Q_h})}$.

The eigenfrequencies ($\omega$) of the Helmholtz
equation (\ref{a5}) are obtained from
the zeros of the determinant of (\ref{gri}).
In the coordinate system $\omega$ versus ${\bf k}_B$,
the photonic band diagrams are given by the trajectories of the
eigenfrequencies $\omega$
when ${\bf k}_B$ follows the boundary of one of the
irreducible regions in the first Brillouin zone
(see Fig.~\ref{IBZ}).

In the case of metallic media, ${\bf k}_B$
becomes complex. It is then simpler to use a
summation method \cite{jmpls} different from (\ref{rls}),
and a technique, based on Bloch's theorem,
developed originally by McRae\cite{mcrae} in low energy electron
diffraction, and applied recently to photonic crystals by
Gralak {\em et al.} \cite{gralak}.
We consider the array as an infinite
stack of gratings and calculate the scattering
matrices for a single grating in the array \cite{josa12}.
Then, the complex
crystal momentum ${\bf k}_B$ is found by solving an eigenvalue problem
\cite{josa3}.

\section{The Long Wavelength Limit}
\label{longwavelentgh}

\subsection{Partial Resonances}
\label{prlwl}

The solution of the problem of electromagnetic scattering by
a solid cylinder, at normal incidence and in the long wavelength
limit\cite{ray2}, revealed the occurrence of
a resonance in scattering when the relative dielectric
constant of the cylinder was -1. Later, the existence of this resonance
was proved experimentally in studies of radio reflection by meteor
trails\cite{herlofson}, and of plasma columns\cite{tonks,romell}.
Also, Wait\cite{wait1,wait2} has shown that in the long wavelength
limit the scattering resonances given by the formula
\begin{equation}
\left( \mu_1 + \mu_2 \right)
\left( \varepsilon_1 + \varepsilon_2 \right) = 0,
\label{wait}
\end{equation}
where the label 1 refers to the cylinder and 2 refers to the
homogeneous medium outside the cylinder, are independent of the
polarization and angle of incidence.
Note that, in electrostatics the condition
$(\varepsilon_1 + \varepsilon_2)=0$ defines the accumulation point
for the poles and zeros of the effective dielectric constant of a
two--phase composite\cite{bergman,ross},
with $(\mu_1 + \mu_2)=0$ corresponding to magnetostatics.

We applied the method developed by Wait\cite{wait1} to solve
the problem of scattering of a plane wave by a coated cylinder at
oblique incidence. In this case, in the long wavelength limit we
obtained the resonance condition
\begin{equation}
\left( \mu_c + \mu_s \right)
\left( \mu_s + \mu_m \right)
\left( \varepsilon_c + \varepsilon_s \right)
\left( \varepsilon_s + \varepsilon_m \right)
= 0,
\label{us}
\end{equation}
which describes precisely the electrostatic partial resonances of
a three--phase composite\cite{cc1,cc2,cc3}, plus their magnetostatic
counterpart.
The resonance condition (\ref{us}) also suggests a generalization
to a $n$--phase composite, of the form
\begin{equation}
\prod_{i=0}^{n-1}
\left( \mu_i + \mu_{i+1} \right)
\left( \varepsilon_i + \varepsilon_{i+1} \right) = 0,
\end{equation}
for the case of a multi-coated cylinder.
Here, $\mu_{n}$ and $\varepsilon_{n}$ denote the magnetic permeability
and dielectric constant of the innermost phase (core), while
$\mu_{0}$ and $\varepsilon_{0}$ represent the magnetic permeability
and dielectric constant of the medium outside the composite.

\subsection{TE Polarization}
\label{tepr}

We concentrate now on non-magnetic materials for which
$\mu_m=\mu_s=\mu_c=\mu_0$, so that
$\varepsilon_m=n_m^2 \varepsilon_0$,
$\varepsilon_s=n_s^2 \varepsilon_0$ and
$\varepsilon_c=n_c^2 \varepsilon_0$, where $\varepsilon_0$
is the dielectric constant of free space, and $n_i$
($i = m, s, c$) represent the refractive indexes of the matrix, shell
and core, respectively.
In the long wavelength limit ($k_m\rightarrow 0$),
we use the expressions of the
Bessel functions for small arguments\cite{AS} in
the boundary conditions coefficients $M_l$ from (\ref{a21}),
for TE polarization.
For $l\neq0$ we obtain
\begin{equation}
M_l = M_{-l} =
- \frac{1}{\pi} \left( \frac{2}{k_m r_s} \right)^{2 l}
l! (l-1)! \, \gamma_l \, ,
\label{eq6ad}
\end{equation}
where
\begin{equation}
\gamma_l =
\frac{r_c^{2l} (\varepsilon_s - \varepsilon_c)
(\varepsilon_m - \varepsilon_s) + r_s^{2l}
(\varepsilon_s + \varepsilon_c)
(\varepsilon_m + \varepsilon_s)}
{r_c^{2l} (\varepsilon_s - \varepsilon_c)
(\varepsilon_m + \varepsilon_s) + r_s^{2l}
(\varepsilon_s + \varepsilon_c)
(\varepsilon_m - \varepsilon_s)} \, ,
\label{eq6a}
\end{equation}
while for $l=0$, we obtain a completely different form
\begin{equation}
M_0 = - \frac{1}{\pi} \left( \frac{2}{k_m r_s} \right)^4
\frac{\varepsilon_m}{(\varepsilon_s-\varepsilon_c)
(r_c/r_s)^4 + (\varepsilon_m-\varepsilon_s)} \, .
\label{tem0}
\end{equation}

To relate the long wavelength limit of the dynamic problem with
the corresponding problem in electrostatics we apply the same
method as in Ref.~\cite{jewatem}. Thus, the boundary conditions
(\ref{a15x}-\ref{a16}) correspond to an electrostatic problem in which
the inverse of the dielectric constants
($\varepsilon_c\rightarrow 1/\varepsilon_c$,
$\varepsilon_s\rightarrow 1/\varepsilon_s$ and
$\varepsilon_m\rightarrow 1/\varepsilon_m$) have to be considered.
This will also change
$\gamma_l\rightarrow -\gamma_l$. Now, the boundary conditions
(\ref{a21}) for $l\neq0$ can be written in the form
\begin{equation}
A_l^m = - \frac{1}{\pi} \left( \frac{2}{k_m} \right)^{2 l}
l! (l-1)! \, \gamma_l \,
\frac{B_l^m}{r_s^{2l}} .
\label{tepr1}
\end{equation}

In electrostatics, the corresponding relationship between the
coefficients $A_l$ and
$B_l$ which controls the response of a coated cylinder to an
external field, has the form\cite{cc1}
\begin{equation}
A_l' = \gamma_l \, \frac{B_l'}{r_s^{2l}} \, .
\label{statics}
\end{equation}
Note that, due to the symmetry properties of the electrostatic
potential, $l$ cannot
take even values in the case of a square array, or values which are
multiples of 3 in the case of a hexagonal array\cite{perrins}.
By comparing (\ref{tepr1}) with (\ref{statics}) we may
deduce the relation between static and dynamic multipole coefficients
\begin{equation}
B_l' = - \frac{1}{\pi} \left( \frac{2}{k_m} \right)^{2 l}
l! (l-1)! \, B_l^m \, .
\label{tepr2}
\end{equation}
for $l\neq0$.
The dynamic field in the matrix ($H_z^{(m)}$) may be approximated
by\cite{jewatem,siam}
\begin{equation}
H_z^{(m)}({\bf r}) = A_0 \exp{[ i k_B V_m({\bf r}) ]}
+ {\mathcal O}(k_B^2) \, ,
\label{tepr3}
\end{equation}
where $V_m$ is the solution of the static problem.
Note that the derivation of (\ref{tepr3}) does not involve the
boundary conditions coefficient $M_0$ from (\ref{tem0}).

In electrostatics, the partial resonances of a three--phase
composite consisting of coated cylinders are defined by the
equations
\begin{eqnarray}
\varepsilon_c + \varepsilon_s & = & 0 \qquad ({\rm core--shell~
 resonance}),
\label{cspr} \\
\varepsilon_s + \varepsilon_m & = & 0 \qquad ({\rm shell--matrix~
 resonance}),
\label{smpr}
\end{eqnarray}
when (\ref{statics}) becomes
\begin{equation}
A_l' = \frac{\varepsilon_m + \varepsilon_c}{\varepsilon_m -
\varepsilon_c} \, \frac{B_l'}{r_s^{2l}} \, ,
\label{cspres}
\end{equation}
or
\begin{equation}
A_l' = \frac{\varepsilon_m + \varepsilon_c}{\varepsilon_m -
\varepsilon_c} \, \frac{B_l'}{(r_s^2/r_c)^{2l}} \, ,
\label{smpres}
\end{equation}
respectively.
In the first case, the field inside the coated cylinder is exactly
the same as would be found within a solid cylinder of radius $r_s$
and dielectric constant $\varepsilon_c$, while the potential
outside the coated cylinder, in the matrix, is precisely the same
as that outside the solid cylinder\cite{cc1}.
The second case corresponds to an effective dielectric constant of
the array of coated cylinders, identical with that of an array of
solid cylinders of radius $r_s^2/r_c > r_s$ (the geometrical image of
the core boundary with respect to the shell outer boundary), and
dielectric constant $\varepsilon_c$. Now, the field external to the coated
cylinder {\em and beyond the radius} $r_s^2/r_c$ is the same as that
external to the solid cylinder\cite{cc1}.

Since it is the relationship between the coefficients $A_l$ and
$B_l$ which controls the response of a coated cylinder to an
external field,
equations (\ref{statics}) and (\ref{tepr1}) show that this
response is determined by $\gamma_l$ in electrostatics as well
as in the long wavelengths limit of electrodynamics.
The limiting process is smooth and,
therefore, we expect a resonant behaviour of the array of
coated cylinders, for nonzero frequencies, when one of the
conditions (\ref{cspr}) or (\ref{smpr}) is satisfied.

\subsection{TM Polarization}
\label{tmpr}

The behaviour of the array of coated cylinders is completely
different in the case of TM polarization.
Now, in the long wavelength limit,
the boundary conditions coefficients $M_l$ from (\ref{a21})
take the form
\begin{eqnarray}
&& M_l  =  M_{-l} \nonumber \\
&& =
\frac{1}{\pi} \left( \frac{2}{k_m r_s} \right)^{2 l + 2}
\frac{2 \, l! (l+1)! \, \varepsilon_m}{(\varepsilon_c - \varepsilon_s)
(r_c/r_s)^{2l+2} + (\varepsilon_s - \varepsilon_m)}  ,
\label{tmpr1}
\end{eqnarray}
for $l\neq0$, and
\begin{equation}
M_0 =
\frac{1}{\pi} \left( \frac{2}{k_m r_s} \right)^{2}
\frac{\varepsilon_m}{(\varepsilon_c - \varepsilon_s)
(r_c/r_s)^{2} + (\varepsilon_s - \varepsilon_m)} \, .
\label{tmpr2}
\end{equation}
Following the same method as in Ref.\cite{jewatem} we consider the
central equation ($l=0$) of the Rayleigh identity (\ref{gri})
\begin{equation}
\left[ M_0 + S_0^Y(k_m, {\bf k}_0) \right] B_0 + \sum_{n\neq0}
S_{-n}^Y(k_m, {\bf k}_0) B_n = 0 \, ,
\label{gri0}
\end{equation}
and employ the forms of lattice sums for small
$k_m$\cite{jewatem}
\begin{eqnarray}
S_0^Y & \sim & \frac{4}{A k_0^2 (\alpha^2 -1)} \, , \\
S_1^Y & \sim & \frac{4 i}{A k_0^2 \alpha (\alpha^2 -1)}
\, e^{i \theta_0} \, , \\
S_n^Y & \sim & {\mathcal O}(k_0^{-n}) \, , \quad {\rm for}~l\geq2
\, , \quad
\left( S_{-n}^Y = \overline{S_n^Y} \right)  ,
\end{eqnarray}
where the superposed bar denotes complex conjugation,
$A$ is the area of the Wigner-Seitz cell,
$\theta_0=\arg{({\bf k}_0)}$ and $\alpha=k_m/k_0$.
We also introduce the relations between dynamic ($B_n$) and static
($B_n'$) multipole coefficients\cite{jewatem}
\begin{eqnarray}
B_n & \sim & (k_m r_s)^{n+2} \, B_n' \, , \\
B_{-n} & \sim & (k_m r_s)^{n+2} \, \overline{B_n'} \, ,
\end{eqnarray}
for $n\geq0$. Here, the static multipole
coefficients are, to leading order, independent of $k_m$.
Finally, the equation (\ref{gri0}) becomes
\begin{eqnarray}
\left[
\frac{\varepsilon_m}{(\varepsilon_c - \varepsilon_s)
(r_c/r_s)^{2} + (\varepsilon_s - \varepsilon_m)}
- \frac{\pi \alpha^2 r_s^2}{A (1 - \alpha^2)} \right] B_0'
\nonumber \\
+ \sum_{n\neq0} B_n' \, {\mathcal O}(k_m) = 0 .
\label{tmpr3}
\end{eqnarray}
From all the other equations ($l\neq0$) of (\ref{gri}) we obtain
$B_l'=0$ so that, in the long wavelengths limit,
to leading order, the only non-trivial solution
of the Rayleigh identity (\ref{gri}) is
$B_0'\neq0$ and $B_n'=0$ ($\forall n\neq0$), if
\begin{equation}
\frac{\varepsilon_m}{\alpha^2} =
f_c \varepsilon_c + f_s \varepsilon_s + f_m \varepsilon_m \, ,
\label{tmpr4}
\end{equation}
with
$f_c=\pi r_c^2/A$, $f_s=\pi(r_s^2-r_c^2)/A$ and $f_m=1-\pi
r_s^2/A$,
denoting the area fractions of the phases inside the unit cell.
From (\ref{fields}) we derive the form of the electric
field in the matrix:
\begin{equation}
E_z^{(m)} \sim \frac{\varepsilon_m}{(\varepsilon_c - \varepsilon_s)
(r_c/r_s)^{2} + (\varepsilon_s - \varepsilon_m)} B_0' \, .
\label{tmpr5}
\end{equation}

Equation (\ref{tmpr4}) represents the linear mixing formula
\cite{wiener}
for the effective dielectric constant of an array of coated
cylinders subject to a uniform field applied along the cylinder
axis. Note that
there are no terms of the form $\varepsilon_c + \varepsilon_s$ or
$\varepsilon_s + \varepsilon_m$ in (\ref{tmpr3}) or (\ref{tmpr4})
to indicate a core--shell or shell--matrix partial resonance.
Again, the limiting process is smooth and,
consequently, we do not expect a resonant behaviour of the array of
coated cylinders, for any frequency, in the case of TM polarization.

\section{Numerical Results}
\label{numerics}

In this section we discuss the behaviour of a three--phase
composite consisting of aluminium oxide (\alo), whose refractive index
depends on the frequency, and two pure dielectric phases:
one with a refractive index $n_d=1.37$, and vacuum ($n_0=1$).

%%%%%%%%%%%%%%%%%%%%%%%%%%%%%%%%%%%%%%%%%%%%%%%%%%%%%%%%%%%%%%%%%%%%%%%
\begin{figure}[h]
\vspace{1ex}
\centerline{\psfig{file=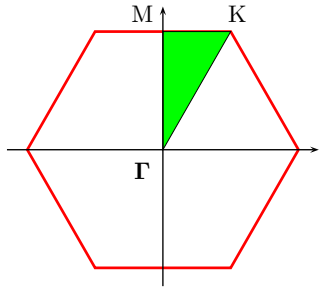,height=1.5in}}
\vspace{3ex}
\caption{ The irreducible part (green triangle) of the first
Brillouin zone.} \label{IBZ}
\end{figure}
%%%%%%%%%%%%%%%%%%%%%%%%%%%%%%%%%%%%%%%%%%%%%%%%%%%%%%%%%%%%%%%%%%%%%%%

Note that, all the defining equations for partial resonances
and the boundary conditions coefficients, in the long wavelengths
limit or in electrostatics, are homogeneous functions of
dielectric constants.
Also, note that the phases are non--magnetic materials.
Hence, in this section we will denote by
$\varepsilon$ the relative dielectric constant
$\varepsilon/\varepsilon_0=n^2$ (the square of the refractive index).
With this notation, we use
the experimental data for the real and imaginary parts
of the refractive index of aluminium oxide in the region
$0.3\mu{\rm m}\le\lambda\le20.0\mu{\rm m}$\cite{palik},
to evaluate the relative dielectric constant of aluminium oxide
($\varepsilon_{\alo}=n_{\alo}^2$).
Also, the relative dielectric constant of the second phase is
$\varepsilon_d=1.876$.

We assume that
the composite is structured as a hexagonal array
(array constant $d$=1$\mu$m) of coated
cylinders (core radius $r_c=0.03\mu$m,
cylinder radius (core+coating) $r_s=0.31\mu$m)
in vacuum.
The Wigner--Seitz cell is shown in Fig.~\ref{WSC}, while
the irreducible part of the first Brillouin zone (K$\Gamma$MK)
is shown Fig.~\ref{IBZ}.
For this region of the first Brillouin zone, the coordinates of the
corners are
$${\rm K} \left( \frac{2\pi}{3 d}, \frac{2\pi}{d \sqrt{3}} \right),
\quad \Gamma (0, 0) ,
\quad {\rm M} \left( 0, \frac{2\pi}{d \sqrt{3}} \right).$$

\subsection{Partial Resonances}
\label{pr}

In the case of an absorbing (metallic--type) core and
a dielectric shell,
the constant, the positive value of the
shell dielectric constant
($\varepsilon_s=n_d^2$)
makes impossible a relation of the type
$n_d^2+n_m^2=0$ corresponding to a shell--matrix
partial resonance.
Hence, we expect only core--shell partial resonances, which are
defined by the equation
\begin{equation}
n_{\alo}^2 + n_d^2 =0 \, .
\label{cs01}
\end{equation}

%%%%%%%%%%%%%%%%%%%%%%%%%%%%%%%%%%%%%%%%%%%%%%%%%%%%%%%%%%%%%%%%%%%%%%%
\begin{figure}[h]
\centerline{\psfig{file=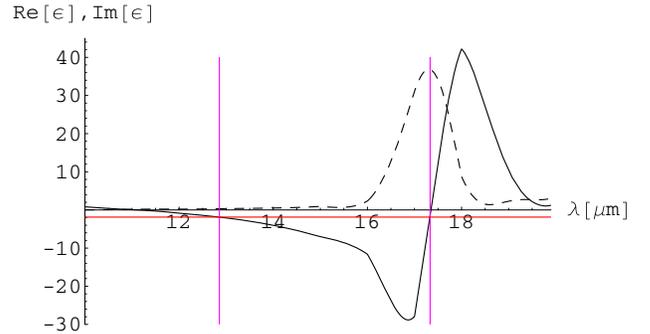,width=3.3in}}
\vspace{2ex}
\caption{Real (solid curve) and imaginary (dashed curve) part of
$\alo$ relative dielectric constant versus wavelength. The
horizontal red line shows $-n_d^2$, while the vertical magenta
lines mark the two solutions $\lambda_1$ and $\lambda_2$, of
(\protect\ref{realcs01}).} \label{al2o3eps}
\end{figure}
%%%%%%%%%%%%%%%%%%%%%%%%%%%%%%%%%%%%%%%%%%%%%%%%%%%%%%%%%%%%%%%%%%%%%%%

Note that, $n_d^2$ is real and so we will solve
the real part of (\ref{cs01}):
\begin{equation}
\Re e \left( n_{\alo}^2 + n_d^2 \right) =0 \, ,
\label{realcs01}
\end{equation}
and then check the magnitude of the imaginary part.

In the wavelengths domain $0.3\mu{\rm m}\le\lambda\le20.0\mu{\rm
m}$, Eq.~(\ref{realcs01}) has two solutions
\begin{equation}
\lambda_1 \approx 12.861\mu{\rm m},
%\qquad
%\omega_1 d/(2 \pi c) \approx 0.0778,
\label{pr01}
\end{equation}
giving a value of
\begin{equation}
n_{\alo}^2 + n_d^2 \approx 10^{-5} + 0.329\, i \, ,
\label{pr02}
\end{equation}
when substituted in (\ref{cs01}), and
\begin{equation}
\lambda_2 \approx 17.333\mu{\rm m},
%\qquad
%\omega_2 d/(2 \pi c) \approx 0.0577,
\label{pr03}
\end{equation}
giving a value of
\begin{equation}
n_{\alo}^2 + n_d^2 \approx 10^{-4} + 32.72\, i \, ,
\label{pr04}
\end{equation}
when substituted in (\ref{cs01}). They are marked by the magenta
vertical lines in Fig.~\ref{al2o3eps}. Actually, the magnitude of
the imaginary part of (\ref{cs01}) shows that we may have a
partial core--shell resonance at $\lambda_1$, while the second
root appears only as the result of an anomaly of $n_{\alo}^2$
around $\lambda_2$.

In the case of dielectric core and absorbing (metallic--type)
shell we expect the coated cylinders to exhibit both types of
partial resonances: core--shell $$n_d^2+n_{\alo}^2=0$$ and
shell--matrix $$n_{\alo}^2 + n_m^2=0.$$
The wavelengths at which the first type of partial resonances
occurs are given by Eq.~(\ref{realcs01}), while
the second type, is identical with the classical
resonance\cite{bergman,ross} exhibited by solid cylinders
from $\alo$ in vacuum.

%%%%%%%%%%%%%%%%%%%%%%%%%%%%%%%%%%%%%%%%%%%%%%%%%%%%%%%%%%%%%%%%%%%%%%%
\begin{figure}[h]
\vspace{2ex}
\centerline{\psfig{file=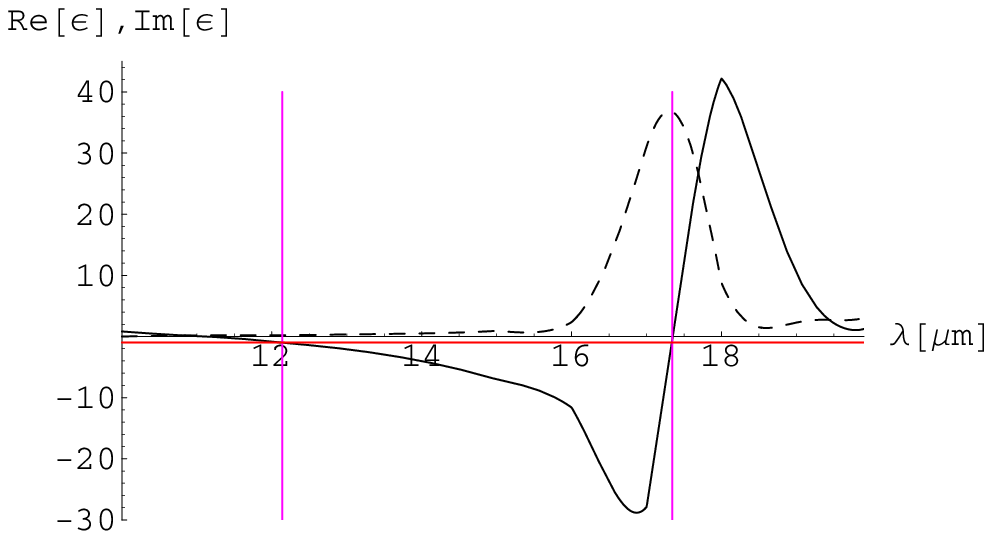,width=3.3in}}
\vspace{3ex}
\caption{Real (solid curve) and imaginary (dashed curve) part of
$\alo$ relative dielectric constant versus wavelength. The
horizontal red line shows $-n_m^2=-1$, while the vertical magenta
lines mark the two solutions $\widetilde{\lambda}_1$,
$\widetilde{\lambda}_2$, of (\protect\ref{realsm01}).}
\label{al2o3eps1}
\end{figure}
%%%%%%%%%%%%%%%%%%%%%%%%%%%%%%%%%%%%%%%%%%%%%%%%%%%%%%%%%%%%%%%%%%%%%%%

Thus, for the shell--matrix resonances, in the range
$0.3\mu{\rm m}\le\lambda\le20.0\mu{\rm m}$,
the equation
\begin{equation}
\Re e \left( n_{\alo}^2 + n_m^2 \right) =0 \, ,
\label{realsm01}
\end{equation}
has two solutions (see Fig.~\ref{al2o3eps1}):
\begin{equation}
\widetilde{\lambda}_1 \approx 12.151\mu{\rm m},
%\qquad
%\widetilde{\omega}_1 d/(2 \pi c) \approx 0.0823,
\label{pr05}
\end{equation}
giving a value of
\begin{equation}
n_{\alo}^2 + n_m^2 \approx 10^{-15} + 0.234\, i \, ,
\label{pr06}
\end{equation}
and
\begin{equation}
\widetilde{\lambda}_2 \approx 17.344\mu{\rm m},
%\qquad
%\widetilde{\omega}_2 d/(2 \pi c) \approx 0.0577,
\label{pr07}
\end{equation}
giving a value of
\begin{equation}
n_{Al_2O_3}^2 + n_m^2 \approx 10^{-13} + 32.611\, i \, .
\label{pr08}
\end{equation}
These wavelengths are marked by magenta lines in
Fig.~\ref{al2o3eps1}. Again, the magnitude of the imaginary part
of (\ref{pr06}) and (\ref{pr08}) shows that we may have a partial
core--shell resonance at $\widetilde{\lambda}_1$, while the second
root appears only as the result of the anomaly of $n_{\alo}^2$
around $\widetilde{\lambda}_2$.

%%%%%%%%%%%%%%%%%%%%%%%%%%%%%%%%%%%%%%%%%%%%%%%%%%%%%%%%%%%%%%%%%%%%%%%
\begin{figure}[h]
\vspace{3ex}
\centerline{\psfig{file=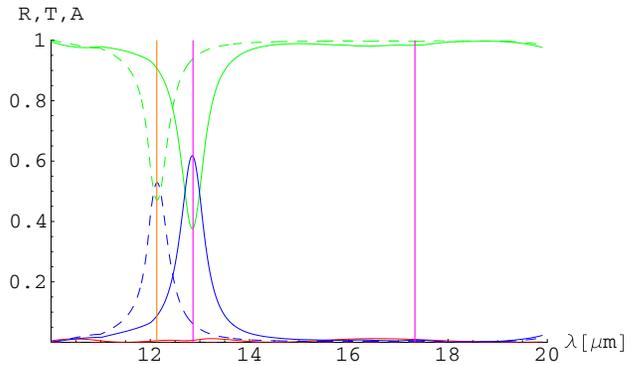,width=3.3in}}
\vspace{3ex}
\caption{TE polarization: Reflectance (red),
transmittance (green) and absorptance (blue) for a stack of solid
cylinders of refractive index $n_c=n_{\alo}$ and radius
$r_c=0.03\mu$m (dashed curves), and for a stack of coated
cylinders (solid curves). In both cases the stack consists of 30
gratings of cylinders, and represents a ``slice'' from a
2-dimensional hexagonal array. The vertical magenta lines mark the
two solutions $\lambda_1$ and $\lambda_2$, of
(\protect\ref{realcs01}).}
\label{rtateccco}
\end{figure}
%%%%%%%%%%%%%%%%%%%%%%%%%%%%%%%%%%%%%%%%%%%%%%%%%%%%%%%%%%%%%%%%%%%%%%%

\subsection{Absorbing Core and Dielectric Shell}
\label{mcdsresults}

Here, we consider that the coated cylinders have
an aluminium oxide core ($n_c=n_{\alo}$),
a dielectric coating ($n_s=n_d$), and are embedded in vacuum ($n_m=1$).
The coating is a lossless, dielectric material of
constant (i.e., independent of frequency)
refractive index $n_s$=1.37.
This is essentially the same as the refractive index
of MgF$_2$ for
$1\mu{\rm m} \lesssim \lambda \lesssim 2.5\mu{\rm m}$\cite{palik,duncanson}.

\subsubsection{TE Polarization}
\label{teres}

In Fig.~\ref{rtateccco} we show
a comparison between two types of stacks of gratings:
a stack of 30 gratings of solid cylinders having a refractive index
$n_c=n_{\alo}$ and radius $r_c=0.03\mu{\rm m}$ (dashed curves),
and for a stack of 30 gratings of coated cylinders defined
at the beginning of Sec.~\ref{mcdsresults} (solid curves).
In both cases, the period of the gratings is $d=1\mu{\rm m}$.
Actually, we compare the stack of coated cylinders with a stack
in which the cylinder coatings have been removed (but keeping al the
other parameters unchanged), to show the
influence of the shell on the optical properties of the stack
(reflectance, transmittance and absorptance), for normal incidence
and TE polarization.

\begin{figure}[h]
\centerline{\psfig{file=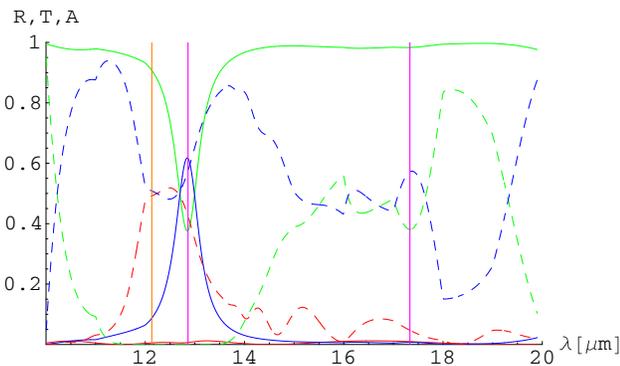,width=3.3in}}
\vspace{3ex}
\caption{TE polarization: Reflectance (red),
transmittance (green) and absorptance (blue) for a stack of solid
cylinders of refractive index $n_c=n_{\alo}$ and radius
$r_s=0.31\mu$m (dashed curves), and for a stack of coated
cylinders (solid curves). In both cases the stack consists of 30
parallel gratings of cylinders, and represents a ``slice'' from a
2-dimensional hexagonal array. The vertical magenta lines mark the
two solutions $\lambda_1$ and $\lambda_2$, of
(\protect\ref{realcs01}).}
\label{rtatccecdetail}
\end{figure}

We can see in Fig.~\ref{rtateccco} an enhancement in absorptance
for the stack of coated cylinders at $\lambda_1$, while the stack
of solid cylinders exhibits an anomalous absorptance at $\lambda_c
\approx 12.151\mu{\rm m}<\lambda_1$, which corresponds to the
classical resonance of a two--phase composite
$n_{\alo}^2+n_m^2=0$\cite{bergman,ross} (in fact, the wavelength
for this classical resonance, $\lambda_c=\widetilde{\lambda}_1$,
is given by (\ref{realsm01})) Thus, in the case of TE polarization
there is a partial resonance of the type core--shell at $\lambda_1
= 12.861\mu{\rm m}$. At this wavelength, the anomalous increase of
the absorptance of coated cylinders shows a lossy (or metallic)
type behaviour, which is a characteristic of aluminium oxide and
is never exhibited by the dielectric shell (lossless media). Note
that the core radius is ten times smaller than the shell radius.
We can also see in Fig.~\ref{rtatccecdetail} that,  at $\lambda_1$
the absorptance of the stack of coated cylinders (blue solid
curve) is approximatively equal with the absorptance of a stack of
solid cylinders, made from Al$_2$O$_3$, and having the radius
$r_s=0.31\mu$m (blue dashed curve).

%%%%%%%%%%%%%%%%%%%%%%%%%%%%%%%%%%%%%%%%%%%%%%%%%%%%%%%%%%%%%%%%%%%%%%%
%%%%%%%%%%%%%%%%%%%%%%%%%%%%%%%%%%%%%%%%%%%%%%%%%%%%%%%%%%%%%%%%%%%%%%%

Note the oscillations of the absorptance and transmittance, for
the stack of solid cylinders of radius $r_s$ and refractive index
$n_c$ (Fig.~\ref{rtatccecdetail}). In contrast with this
behaviour, the absorptance and the transmittance of the stack of
coated cylinders are almost constant, except a small region around
$\lambda_1$. If we label by $cc$ the reflectance ($R_{cc}$),
transmittance ($T_{cc}$) and absorptance ($A_{cc}$) for the stack
of coated cylinders, and by $sc$ the corresponding quantities
($R_{sc}$, $T_{sc}$, $A_{sc}$) for the stack of solid cylinders,
then, in the neighbourhood of the core--shell partial resonance,
we have:
\begin{equation}
R_{cc} \approx T_{sc} \approx 0, \quad
T_{cc} \approx R_{sc}, \quad
A_{cc} \approx A_{sc} .
\label{scRTA}
\end{equation}
Equation (\ref{scRTA}) shows that the {\em equivalence} between the
two types of stacks is actually a sort of {\em duality}in the sense that
they have similar absorptance, but the resonant system resembles a
material which absorbs and transmits the incident radiation, while
the stack of \alo cylinders absorbs and reflects incident
radiation.

In conclusion, at $\lambda_1$, for normal incidence and TE
polarization, the stack of coated cylinders with a very small
metallic--type core, has really the same absorptance as a stack of
solid metallic cylinders of an enlarged radius $r_s=0.31\mu$m. It
looks like the core and the shell {\em are replaced} by a new
material which has the same absorptance as the core material, an
effect identical with the core--shell partial resonance from
electrostatics\cite{cc1,cc2,cc3}.

%%%%%%%%%%%%%%%%%%%%%%%%%%%%%%%%%%%%%%%%%%%%%%%%%%%%%%%%%%%%%%%%%%%%%%%
\begin{figure}[h]
\vspace{1ex}
\centerline{\psfig{file=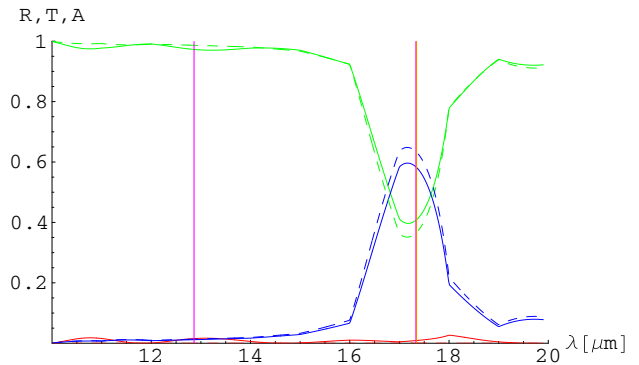,width=3.3in}}
\vspace{3ex}
\caption{TM polarization: Reflectance (red),
transmittance (green) and absorptance (blue) for a stack of solid
cylinders of refractive index $n_c=n_{\alo}$ and radius $r_c$
(dashed curves), and for a stack of coated cylinders (solid
curves). In both cases the stack consists of 30 gratings of
cylinders, and represents a ``slice'' from a 2-dimensional
hexagonal array. The vertical magenta lines mark the two solutions
$\lambda_1$ and $\lambda_2$, of (\protect\ref{realcs01}).}
\label{rtatmccco}
\end{figure}
%%%%%%%%%%%%%%%%%%%%%%%%%%%%%%%%%%%%%%%%%%%%%%%%%%%%%%%%%%%%%%%%%%%%%%%

\subsubsection{TM Polarization}
\label{tmres}

There is no partial resonance for a stack of coated cylinders, for
TM polarization.
Fig.~\ref{rtatmccco} shows no change of optical characteristics
around $\lambda_1$. The second solution of
(\ref{realcs01}) is $\lambda_2 \approx 17.333\mu{\rm m}$, and at
this wavelength the stack of 30 gratings of coated cylinders
exhibits an anomalous absorptance (see Fig.~\ref{rtatmccco}),
which corresponds to the large imaginary part of (\ref{pr04}). The
stack of solid cylinders having a refractive index
$n_c=n_{\alo}$ and radius $r_c=0.03\mu{\rm m}$, exhibits an
anomalous absorptance (see Fig.~\ref{rtatmccco}) at
$\widetilde{\lambda}_2$, corresponding to (\ref{pr08}).

In both cases, the anomalous high absorptance is not a resonance
effect, and it is produced by an anomalous behaviour of the
relative dielectric constant of $\alo$. At
$\lambda_2\approx\widetilde{\lambda}_2$ the real part of the
relative dielectric constant of $\alo$ is relatively small, while
the imaginary part reaches a maximum value of $\approx 30$ (see
Fig.~\ref{al2o3eps}). Therefore the absorptance of the core of
coated cylinders, or of the thin cylinders of core material only
becomes large (see Fig.~\ref{rtatmccco}). At the same time, for
long wavelengths and TM polarization, the effective dielectric
constant of an array of cylinders is given by the linear mixing
formula (\ref{tmpr4}). Thus, for an array of coated cylinders the
relative effective dielectric constant is
\begin{equation}
\varepsilon^* =
f_c \frac{\varepsilon_c}{\varepsilon_m} +
f_s \frac{\varepsilon_s}{\varepsilon_m} + f_m \, ,
\label{cclinmix}
\end{equation}
while for the array of solid cylinders of core material and radius
$r_c$ we have
\begin{equation}
\varepsilon^* =
f_c \frac{\varepsilon_c}{\varepsilon_m} + f_m \, .
\label{colinmix}
\end{equation}
Hence, for a pure dielectric shell ($\varepsilon_s=$constant),
in both cases, the behaviour of
$\varepsilon^*$ is determined by $\varepsilon_c$ and this explains
the common maximum in absorptance shown in
Fig.~\ref{rtatmccco}, at the same position as the maximum in the
imaginary part of the relative dielectric constant of $\alo$
(see Fig.~\ref{al2o3eps}).
Also, around the wavelength associated with this maximum, the real part of
the relative dielectric constant of $\alo$ exhibits a sudden
jump from negative to positive values, which makes almost equal
the roots of the equations (\ref{realcs01}) and (\ref{realsm01}).
Actually, the anomalous absorptance exhibited by
stacks of thin, solid cylinders has been described in
Ref.~\cite{anomaly}.

\subsection{Dielectric Core and Absorbing Shell}
\label{dcmsresults}

Here, we consider the complementary structure with coated cylinders, having
a dielectric core ($n_c=n_d$) and
an aluminium oxide coating ($n_s=n_{\alo}$),
embedded in vacuum ($n_m=1$).
Now, the core is a lossless, dielectric material of
constant $\varepsilon_c$.
In this case, we expect both types of partial resonances to occur,
and in both cases the coated cylinder has to exhibit the characteristics
of a solid cylinder made from the core material.
Absorptance is a characteristic feature of the metallic phase.
Therefore, we expect a decrease of the total absorptance for
the stack of gratings,
around $\lambda_1$ and $\widetilde{\lambda}_1$.

\subsubsection{TE Polarization}
\label{teres1}

\begin{figure}[h]
\vspace{2ex}
\centerline{\psfig{file=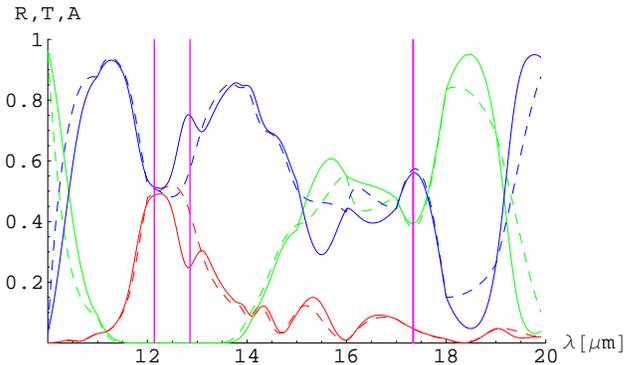,width=3.3in}}
\vspace{2ex}
\caption{TE polarization: Reflectance (red), transmittance (green)
and absorptance (blue) for a stack of solid cylinders of
refractive index $n_c=n_{\alo}$ and radius $r_s=0.31$ (dashed
curves), and for a stack of coated cylinders with $r_c=0.03$ and
$r_s=0.31$ (solid curves). In both cases the stack consists of 30
gratings of cylinders in a hexagonal array. The vertical magenta
lines mark the solutions $\lambda_1$ and $\lambda_2$ of
(\protect\ref{realcs01}), and $\widetilde{\lambda}_1$ and
$\widetilde{\lambda}_2$ of (\protect\ref{realsm01}).}
\label{rtatecc1}
\end{figure}

For TE polarization the reflectance, transmittance and absorptance
of the stack of coated cylinders with dielectric core and metallic
shell, are almost the same as in the case of a stack of solid
cylinders of radius $r_s$, made from aluminium oxide (see
Fig.~\ref{rtatecc1}). Thus, in the neighbourhood of the
shell--matrix resonance at $\widetilde{\lambda}_1 \approx
12.151\mu{\rm m}$ the two stacks have the same optical
characteristics. In fact, Eq.(\ref{smpres}) shows that the coated
cylinders must have optical characteristics similar to those of
solid cylinders made from core material and of an extended radius
$a=r_s^2/r_c$. In our case this means {\em equivalent} dielectric
cylinders of radius $a=3.2\mu{\rm m}>\!\!>d/2=0.5\mu{\rm m}$, i.e.
we have a stack of intersecting cylinders forming a homogeneous,
dielectric slab, and we cannot apply the formulation from
Sec.~\ref{raysec}. In the neighbourhood of the core--shell
resonance at $\lambda_1 \approx 12.861\mu{\rm m}$, the reflectance
of the stack of coated cylinders is diminished, while the
absorptance is increased, compared with a stack of aluminium oxide
cylinders (Fig.~\ref{rtatecc1}). At
$\lambda_2\approx\widetilde{\lambda}_2$ the two types of stacks
exhibit identical optical characteristics, and the local maximum
in absorptance appears due to the anomalous behaviour of
$\varepsilon_{\alo}$ around this wavelength (see
Fig.~\ref{al2o3eps1}).

%%%%%%%%%%%%%%%%%%%%%%%%%%%%%%%%%%%%%%%%%%%%%%%%%%%%%%%%%%%%%%%%%%%%%%%
%%%%%%%%%%%%%%%%%%%%%%%%%%%%%%%%%%%%%%%%%%%%%%%%%%%%%%%%%%%%%%%%%%%%%%%

\begin{figure}[h]
\centerline{\psfig{file=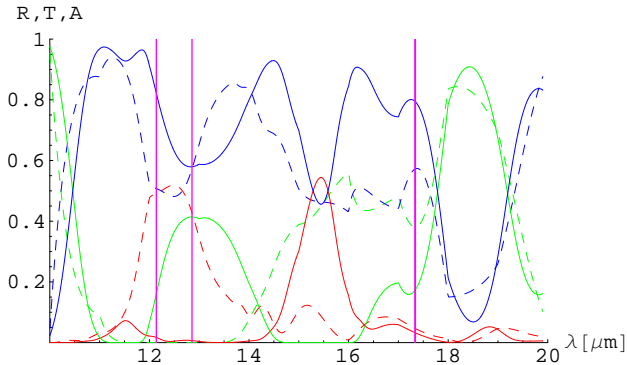,width=3.3in}}
\vspace{2ex}
\caption{TE polarization: Reflectance (red), transmittance (green)
and absorptance (blue) for a stack of solid cylinders of
refractive index $n_c=n_{\alo}$ and radius $r_s=0.31$ (dashed
curves), and for a stack of coated cylinders with $r_c=0.21$ and
$r_s=0.31$ (solid curves). In both cases the stack consists of 30
gratings of cylinders in a hexagonal array. The vertical magenta
lines mark the solutions $\lambda_1$ and $\lambda_2$ of
(\protect\ref{realcs01}), and $\widetilde{\lambda}_1$ and
$\widetilde{\lambda}_2$ of (\protect\ref{realsm01}).}
\label{rtatecc2}
\end{figure}

To avoid such situations when the resonant system cannot be
modelled by the formulation from Sec.~\ref{raysec}, we consider a
stack of coated cylinders with a larger core $r_c=0.21$. In this
case, a shell--matrix partial resonance will lead us to an
equivalent dielectric cylinder, having a radius $a=0.46\mu{\rm
m}<d/2$. Now, in the neighbourhood of the shell--matrix resonance
at $\widetilde{\lambda}_1$, the stack of coated cylinders exhibits
a drop in absorptance, an increased transmittance and a very small
reflectance, while a stack of aluminium oxide cylinders having the
radius $r_s$ has a zero transmittance and equal reflectance and
absorptance (see Fig.~\ref{rtatecc2}). The equivalent stack,
formed by dielectric cylinders of radius $a$, has a zero
absorptance by definition. For long wavelengths ($\lambda>d$) it
has a low reflectance and a high transmittance, therefore, the
only indication of the existence of a shell--matrix partial
resonance is the nonzero transmittance exhibited by the coated
cylinders.

%%%%%%%%%%%%%%%%%%%%%%%%%%%%%%%%%%%%%%%%%%%%%%%%%%%%%%%%%%%%%%%%%%%%%%%
%%%%%%%%%%%%%%%%%%%%%%%%%%%%%%%%%%%%%%%%%%%%%%%%%%%%%%%%%%%%%%%%%%%%%%%

%%%%%%%%%%%%%%%%%%%%%%%%%%%%%%%%%%%%%%%%%%%%%%%%%%%%%%%%%%%%%%%%%%%%%%%
\begin{figure}[h]
\vspace{3ex}
\centerline{\psfig{file=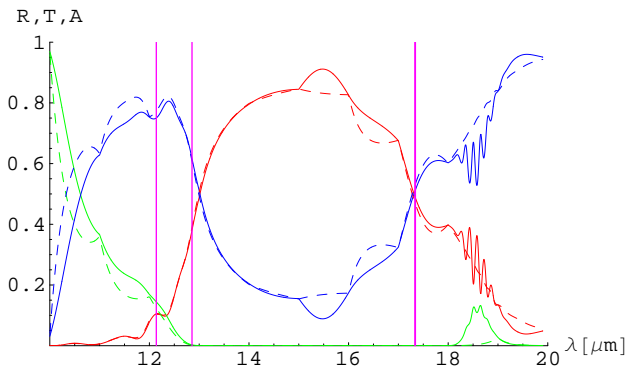,width=3.3in}}
\vspace{2ex}
\caption{TM polarization: Reflectance (red), transmittance (green)
and absorptance (blue) for a stack of solid cylinders of
refractive index $n_c=n_{\alo}$ and radius $r_s=0.31$ (dashed
curves), and for a stack of coated cylinders with $r_c=0.03$ and
$r_s=0.31$ (solid curves). In both cases the stack consists of 30
gratings of cylinders in a hexagonal array. The vertical magenta
lines mark the solutions $\lambda_1$ and $\lambda_2$ of
(\protect\ref{realcs01}), and $\widetilde{\lambda}_1$ and
$\widetilde{\lambda}_2$ of (\protect\ref{realsm01}).}
\label{rtatmcc1}
\end{figure}
%%%%%%%%%%%%%%%%%%%%%%%%%%%%%%%%%%%%%%%%%%%%%%%%%%%%%%%%%%%%%%%%%%%%%%%

In the neighbourhood of the core--shell resonance (at
$\lambda_1$), Fig.~\ref{rtatecc2} shows a reduced absorptance for
the stack of coated cylinders. Again, the equivalent stack,
formed by dielectric cylinders of radius $r_s$, predicted by
(\ref{cspres}), has a zero absorptance by definition, and so the
comparison with a stack of aluminium oxide cylinders of radius $r_s$,
seems to be more natural. In this case, it is interesting to
remark the appearance of the duality mentioned in
Sec.~\ref{teres}.
The two stacks have the same absorptance and the reflectance of
solid cylinders is equal with the transmittance of coated
cylinders
\begin{equation}
R_{cc} \approx T_{sc} \approx 0, \quad
T_{cc} \approx R_{sc}, \quad
A_{cc} \approx A_{sc} .
\label{scRTA1}
\end{equation}

\subsubsection{TM Polarization}
\label{tmres1}

As in the case of TE polarization, when the core radius of the
coated cylinders is small (so that $a=r_s^2/r_c>d/2$),
for TE polarization the reflectance, transmittance and
absorptance of the stack of coated cylinders
with dielectric core and  metallic--type shell, are almost the same
as in the case of a stack of solid cylinders of radius $r_s$,
made from aluminium oxide (see Fig.~\ref{rtatmcc1}).
In fact, there is no resonant behaviour in the whole range
$10\mu{\rm m}<\lambda<20\mu{\rm m}$.

\begin{figure}[h]
\vspace{3ex}
\centerline{\psfig{file=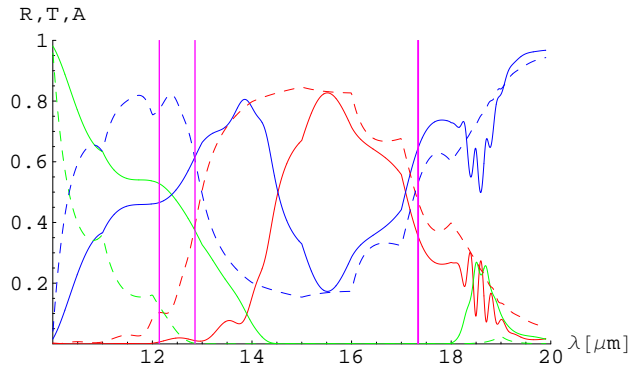,width=3.3in}}
\vspace{3ex}
\caption{TM polarization: Reflectance (red), transmittance (green)
and absorptance (blue) for a stack of solid cylinders of
refractive index $n_c=n_{\alo}$ and radius $r_s=0.31$ (dashed
curves), and for a stack of coated cylinders with $r_c=0.21$ and
$r_s=0.31$ (solid curves). In both cases the stack consists of 30
gratings of cylinders in a hexagonal array. The vertical magenta
lines mark the solutions $\lambda_1$ and $\lambda_2$ of
(\protect\ref{realcs01}), and $\widetilde{\lambda}_1$ and
$\widetilde{\lambda}_2$ of (\protect\ref{realsm01}).}
\label{rtatmcc2}
\end{figure}

In the case of a larger core of coated cylinders, the effect of
shell--matrix partial resonance is small for TM polarization.
Around $\lambda=\widetilde{\lambda}_1$ the absorptance of the
coated cylinders is almost equal with their transmittance, and the
reflectance is close to zero. The curves displayed in
Fig.~\ref{rtatmcc2} show no sudden change around
$\widetilde{\lambda}_1$ (shell--matrix partial resonance) and
$\lambda_1$ (core--shell partial resonance), but it is interesting
to remark the duality (\ref{scRTA1}) at $\lambda_1$.

%%%%%%%%%%%%%%%%%%%%%%%%%%%%%%%%%%%%%%%%%%%%%%%%%%%%%%%%%%%%%%%%%%%%%%%
%%%%%%%%%%%%%%%%%%%%%%%%%%%%%%%%%%%%%%%%%%%%%%%%%%%%%%%%%%%%%%%%%%%%%%%

\section{Conclusions}

We have analyzed the partial resonances of a three--phase
composite, structured as a hexagonal array of coated cylinders, at
long wavelengths. For such structures there are two types of
partial resonances: shell--matrix and core--shell. We have shown
that the shell--matrix resonance has a very small effect. In
contrast, the core--shell resonance appears for TE polarization in
both cases of coated cylinders metal--dielectric--vacuum and
dielectric--metal-- vacuum. Note that, for such resonances the
stack of coated cylinders has a nonzero absorptance, so we cannot
compare it with a stack of dielectric cylinders (in the case of
dielectric--metal--vacuum composite). However, the comparison with
a stack of solid, metallic cylinders of radius $r_s$ shows equal
absorptance and a reflectance--transmittance duality, for the
core--shell resonance.

An interesting behaviour is exhibited by the composite in the
vicinity of shell--matrix ($\widetilde{\lambda}_1$) and
core--shell ($\lambda_1$) resonances (see
Figs.~\ref{rtatccecdetail}, \ref{rtatmccco}, \ref{rtatecc2} and
\ref{rtatmcc2}). In these regions, the composite has a very low
reflectance ($R_{cc} \approx 0$), while the transmittance and
absorptance take similar values.

Generally, a band gap is defined by R=1, and T=0, so that sudden
changes in optical characteristics of an array of coated cylinders
will change the photonic band diagram. In our calculations the
partial resonances occur at long wavelengths (i.e., in the region
of the acoustic band) and we expect the photonic band diagram of
the composite to show horizontal lines intersecting the acoustic
band.
%We are in progress to develop a numerical method
%\cite{josa3} capable to produce the photonic band diagrams for a
%complex crystal momentum ${\bf k}_B$, and thus enabling the study
%of changes in the shape of the acoustic band for multi--phase
%composites containing metallic phases.
These ultrafast changes in the behaviour of a composite,
due to the dependence of the refractive indexes on the wavelength,
can provide the basis of future optical switches in photonic
integrated circuits.

\acknowledgments The Australian Research Council supported this
work. Helpful discussions with Professor J. B. Pendry are acknowledged.

\end{document}